# Thermal bridging of graphene nanosheets via covalent molecular junctions: a Non-Equilibrium Green's Functions - Density Functional Tight-Binding study

Diego Martinez Gutierrez[1], Alessandro di Pierro[1], Alessandro Pecchia[2], Leonardo Medrano Sandonas[3,4], Rafael Gutierrez[3], Mar Bernal[1], Bohayra Mortazavi[5], Gianaurelio Cuniberti[3,4,6], Guido Saracco[1], Alberto Fina[1] (✉)

*1-Dipartimento di Scienza Applicata e Tecnologia, Politecnico di Torino, 15121 Alessandria, Italy*
*2-Consiglio Nazionale delle Ricerche, ISMN, 00017 Monterotondo, Italy*
*3-Institute for Materials Science and Max Bergmann Center of Biomaterials, TU Dresden, 01062 Dresden, Germany*
*4-Center for Advancing Electronics Dresden, TU Dresden, 01062 Dresden, Germany*
*5- Institute of Structural Mechanics, Bauhaus-Universität Weimar, D-99423 Weimar, Germany*
*6- Dresden Center for Computational Materials Science, TU Dresden, 01062 Dresden, Germany.*

**ABSTRACT**

Despite the uniquely high thermal conductivity of graphene is well known, the exploitation of graphene into thermally conductive nanomaterials and devices is limited by the inefficiency of thermal contacts between the individual nanosheets. A fascinating yet experimentally challenging route to enhance thermal conductance at contacts between graphene nanosheets is through molecular junctions, allowing covalently connecting nanosheets otherwise interacting only via weak Van der Waals forces. Beside the bare existence of covalent connections, the choice of molecular structures to be used as thermal junctions should be guided by their vibrational properties, in terms of phonon transfer through the molecular junction. In this paper, density functional tight-binding combined with Green's functions formalism was applied for the calculation of thermal conductance and phonon spectra of several different aliphatic and aromatic molecular junctions between graphene nanosheets. Effect of molecular junction length, conformation, and aromaticity were studied in detail and correlated with phonon tunnelling spectra. The theoretical insight provided by this work can guide future experimental studies to select suitable molecular junctions in order to enhance the thermal transport by suppressing the interfacial thermal resistances, particularly attractive for various systems, including graphene nanopapers and graphene polymer nanocomposites, as well as related devices. In a broader view, the possibility to design molecular junctions to control phonon transport currently appears as an efficient way to produce phononic devices and controlling heat management in nanostructures.

**KEYWORDS**

Thermal conductance, Molecular junctions, Green's functions, DFTB, Graphene, Heat transport, Phonon transmission function.

## 1 Introduction

Heat transfer at interfaces has been recognized as an important issue to control thermal conductivity in nanomaterials, in which most of the physical properties are indeed strongly interface-dependent [1,2]. The thermal resistance associated to each interface is actually representing an obstacle for many technological applications, including heat dissipation in electronics [3,4] and thermal tuning in thermo-electric devices such as Seebeck generators [5-7]. The fundamental phenomena behind the existence of a thermal resistance are related to inefficient phonon transport across the interface. Thus, the study of phonon transport across interfaces is vital for the understanding of the thermal properties in nanomaterials and devices [8], also taking into account that modern electronic devices are experiencing a continuous downsizing towards the nanoscale, thus restraining phonon transport between device boundaries.

Atomistic modelling of phonon transport on nano-devices was done using different methods, ranging from classical elasticity analytic models [9,10] to atomistic molecular dynamics and density functional based methodologies [11,12]. Classical elasticity models provide the fastest solutions, but these require the assumption of a classical continuous media. On the other hand, first-principles density functional models can take into account elaborated details such as the electronic structures in the simulations, but they are computationally demanding and assume that the systems are near the equilibrium state.

In the last decade, a great deal of research interest was focused on graphene, due to its high thermal and electrical conductivity [13-16], optical [17,18] and magnetic [19] properties. In terms of thermal conductivity, graphene was reported both experimentally and theoretically in the range between 1000 and 5000 W/mK [20], depending on the used experimental technique or computational method. Micro-Raman spectroscopy conducted on a suspended single layer graphene gave a thermal conductivity at room temperature up to 5300 W/mK [13]. For graphene, phonons are the prime thermal energy transport channel, as the contribution of phonons to the thermal conductivity is almost 50-100 times greater than that of the electrons [21-23]. Similarly, a *ab-initio* calculations done by Sadeghi et al. [24] showed that the phononic contribution for alkanes is ~ 700 times larger than the electronic contribution. Fano effect, that is the effects of interference between a continuum and a resonant state, is important in the interpretation of electronic transport [25]. Indeed, Famili et al. and Klockner et al. have found that this effect modifies the phonon transport of molecular junctions [26,27]. For phonon transport, the Fano effects occur under Debye

---

energies, for metallic electrodes that is typically in the range of several tens of meV (10 meV≃80.64 cm$^{-1}$)[1], while for graphene, the Debye temperature of $\Theta_D \approx 2100K$ makes that Fano effect limit to 1459.6 $cm^{-1}$ which gives a wide window of phonon energies that contributes to thermal transport at room temperatures [28]. The Debye energy of the metal electrodes sets an upper limit for the energy of the vibrational modes that can contribute to the transport [29]. Furthermore, anharmonic effects might be important at temperatures above the Debye temperature [30]. In fact, for temperatures above the Debye temperature of the graphene sheets, all the phonons of the graphene as well as the vibrational modes of the molecular bridge system with energies within the transport window are thermally occupied, while below this temperature the higher-lying modes are only partially occupied and the thermal conductance hence becomes sensitive to the temperature [31]. Improved coupling of nanoflakes, including graphene, may be obtained by molecular junctions, i.e. an organic molecule covalently bound to both nanoflakes, acting as a thermal bridge between them. While electronic transport in molecular junctions was widely studied in the past decades [32-39], molecular junctions were only recently explored as a tool to tailor the properties of nanophononic devices [24]. Some theoretical studies using molecular dynamics or lattice dynamics have explored phonon transport in molecular junctions. Klockner et al. [31] studied the length dependence of the thermal conductance for single molecule junctions between gold electrodes using density-functional theory (DFT) and nonequilibrium Green's functions (NEGF). Markussen [40] used first principles harmonic approximation and Green's functions to study phonon transport in molecular junctions between silicon nanowires or graphene nanoribbons. Sasikumar et al. [41] used non-equilibrium MD to study phonon transport on different conformations of Si-polyethylene molecules; also by non-equilibrium Molecular Dynamics (MD), Ranganathan et al. [42] studied the possible tuning of thermal conductance by reversible photoisomerism. Li et al. studied the tuning of thermal conductance by mechanical strain [43], and reduction of phonon transport on π-π stacked junctions has also been shown [44]. However, to the best of our knowledge, no systematic studies have been reported so far on the correlation between the molecular structure and the specific transmission spectra between graphene nanosheets through the molecular junction.

In this work, we explored the efficiency in the thermal transport between individual graphene nanosheet through chemical connecting by using a numerical model based on density functional tight binding (DFTB+) [45-48] method combined with Green's functions formalism [49-52], allowing to study coherent phonon transport, thus taking into account elastic scattering across graphene-molecule interfaces as well as along the organic molecules. In particular, several different molecular junctions between graphene sheets with armchair edges were systematically studied in terms of their phononic properties. Hence, the effect of chain length and aromaticity was also studied in details, proving short polyaromatic structures to be the most efficient molecular junction between graphene sheets.

## 2 Simulation Method
### 2.1 Molecular junction models

In this work, we addressed a set of molecular junctions between two graphene sheets. Chemical structures of the junctions (Table 1) were selected to cover aliphatic, aromatic and aliphatic/aromatic structures, with particular focus on those experimentally viable via aryl diazonium edge functionalization of graphene, as recently demonstrated by Bernal et al. [53].

Table 1 Bridging molecules studied in this work. Molecular junctions are designed with various molecules, covalently bound to graphene. For aromatic ends of bridging molecules, grafting to graphene is in para position

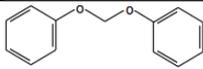

| Aliphatic molecules: –(CH$_2$)$_n$– with $n$ = 1, …, 21 |
|---|

| Aliphatic–aromatic molecules | Aromatic molecules |
|---|---|
| 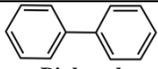 Diphenoxymethane | 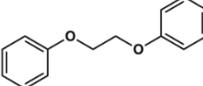 Biphenyl |
| 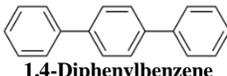 1,2-Diphenoxyethane | 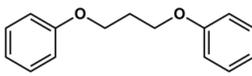 1,4-Diphenylbenzene |
| 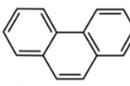 1,3-Diphenoxypropane | 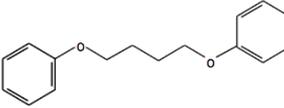 Phenanthrene |
| 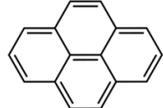 1,4-Diphenoxybutane | 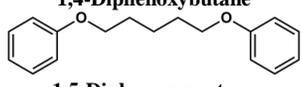 Pyrene |
| 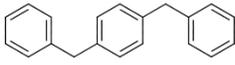 1,5-Diphenoxypentane | 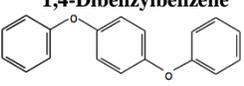 1,4-Dibenzylbenzene |
| 1,4-Diphenoxybenzene | |

Furthermore, elongated vs. compressed conformation of the molecular junction were compared, for selected bridging molecules. The graphene contacts are simulated as semi-infinite by applying periodic boundary conditions in the numerical calculations. Grafting on armchair edges were selected as a model system, for direct comparison with previously reported results on aliphatic junctions [54]. Structural optimization of the nanojunctions were accomplished by using the density functional tight-binding (DFTB) method [45,48,55,56] as implemented in the DFTB+ package [47]. This method combines accuracy with numerical efficiency, and it allows dealing with systems up 2000 atoms in a quantum simulation, especially for carbon-based materials.

The DFTB method is based on a density-functional-based parameterization of a general tight-binding Hamiltonian. Hereby, the system eigenstates are expanded into a linear combination of atomic orbitals (LCAO) minimal valence basis set, which are determined from self-consistent atomic Local Density Approximation (LDA) calculations. Additionally, the effective Kohn-Sham potential is approximated as a superposition of localized potential of neutral atoms (see the supplementary information for more details and a brief summary of the derivation of the method). All nanojunctions were structurally optimized by using steepest descent (SCC calculations [48]) until the absolute value of the interatomic forces was below $10^{-4}$ atomic units with a step size of 3.0 femtoseconds. We have considered periodic boundary conditions along the Y-direction (perpendicular to the transport direction).

---
[1] This relation comes from the energy of the photons E=$\hbar\nu$ ; also 1K≃0.695cm$^{-1}$

## 2.2 Phonon transport properties

In the NEGF method used here, the transport setup is divided into three zones; the central region where the scattering is taking place, and the left and right thermal baths (semi-infinite graphene sheets) as described in Figure 1. The central region not only includes the molecular bridge, but also the part of the pristine graphene sheets directly coupled to the molecules. In this work, the length of pristine graphene was assumed to be ~5 Å, which is long enough as there are no appreciable structural differences in the position of the atoms at that point compared to isolated graphene.

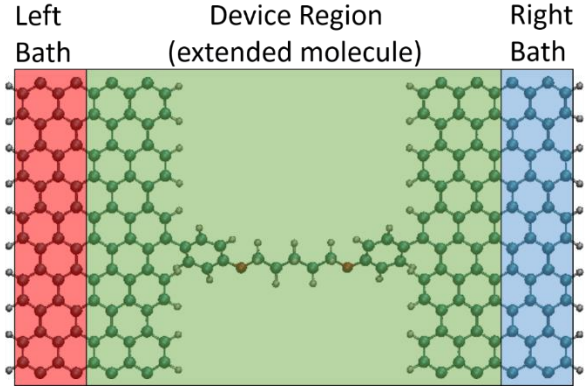

Figure 1 Setup of the different calculation zones for NEGF calculations

The retarded phonon Green's function ($G^r$) for the central region (including the coupling to the thermal baths through self-energies) can be calculated as

$G^r(\omega) = [\omega^2 I - K - \Sigma_L(\omega) - \Sigma_R(\omega)]^{-1}$ (1)

with I is the identity matrix and $\Sigma_{L,R}$ are the self-energies for the left (L) and right (R) baths, respectively, which are calculated as described elsewhere [49,57,58]. Having obtained the Green's Functions and the self-energy matrices, the phonon transmission function, $\tau_{ph}$, can be computed as

$\tau_{ph}(\omega) = Tr[G^r(\omega)\Gamma_L(\omega)G^a(\omega)\Gamma_R(\omega)]$ (2)

Where $\Gamma_{L,R}(\omega) = i[\Sigma^r_{L,R}(\omega) - \Sigma^a_{L,R}(\omega)]$ are the broadening functions and $G^a(\omega) = [G^r(\omega)]^\dagger$ is the advanced Green's function. As far as only elastic processes are considered (no anharmonic interactions or coupling to electronic degrees of freedom), this formalism is mathematically similar to the Landauer approach for charge transport. Thus, for systems, in which elastic scattering is dominant, we can describe phonon transport very accurately [49]. The thermal conductance is defined as

$\kappa_{ph} = \frac{\hbar^2}{2\pi k_B T^2} \int_0^\infty \omega^2 \frac{e^{\hbar\omega/k_B T}}{(e^{\hbar\omega/k_B T}-1)^2} \tau_{ph}(\omega) d\omega$ (3)

Here, $k_B$ and $h$ are the Boltzmann and the Planck constants, respectively. This expression is obtained by a linear expansion in the applied temperature difference $\Delta T$ of the quantity $N_B(T + \Delta T) - N_B(T)$, where $N_B$ is the Bose-Einstein distribution. Moreover, the phonon density of states (DOS) is calculated as

$\eta(\omega) = -\frac{2\omega}{\pi} Tr[Im(G^r(\omega))]$ (4)

These calculations depend on the structure of the system, including the conformation of the molecular junction. Furthermore, the conformation of a molecular junction may evolve in time. However, the calculations made with NEGF involve small oscillations around the equilibrium point. For this reason, several calculations are needed for each system in order to obtain a representative conductance value and phonon spectrum. This is particularly important in the case of un-stretched molecular junctions, for which several different local equilibrium conformation states are possible. By making several calculations of the same system with different initial local equilibrium conformations of the molecular bridge, we were able to obtain the average and the standard deviation of the conductance, the phonon density of states and the transmission function. To obtain representative values of the thermal conductance of the molecular junctions, we considered six different initial conformations which were extracted from molecular dynamic trajectories initialized with different velocities. MD simulations were run for 2 ps by setting an initial temperature of 300 K. Then, these configurations were optimized using DFTB+ until the absolute value of inter-atomic forces was below $10^{-6}$ atomic units. Finally, thermal transport was computed for these conformations using the previously described Green's functions approach.

## 3 Results and discussion

Several different chemical structures were studied as possible candidates for the use as molecular junctions to control thermal conductance between graphene sheets, including aliphatic, aromatic and aliphatic/aromatic molecules (Table 1).

### 3.1 Aliphatic molecular junctions

We start our analysis with the simplest type of molecular bridge, namely simple *n*-alcane chains of varying length. The thermal conductance of such chains as a function of length and temperature is reported in Figure 2a for selected chain lengths (plots for all the chain length n=1 to 21 are reported in Figure S1). These data were obtained from calculating the average over six different random initial conditions on the system. The thermal conductivity dependence on the temperature of these systems has a characteristic shape where the thermal conductance grows with increasing temperature. However, the growth rate decreases with temperature and is in principle expected to reach a plateau at high temperatures. This is theoretically related to the fact that once the maximum number of available vibrational states is occupied (above a certain critical, system-dependent temperature), the integral over the frequency spectrum in Eq. 3 does not change anymore and hence the thermal conductance saturates [31]. Temperatures above 500 K were not addressed in this work as the organic molecular junctions are expected to undergo thermal scission at high temperature, thus making thermal conductance calculation at high temperatures unrepresentative.

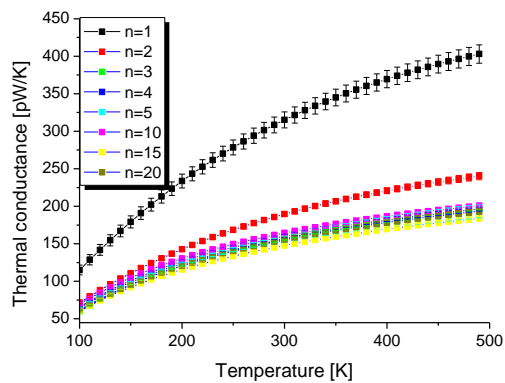

(a)

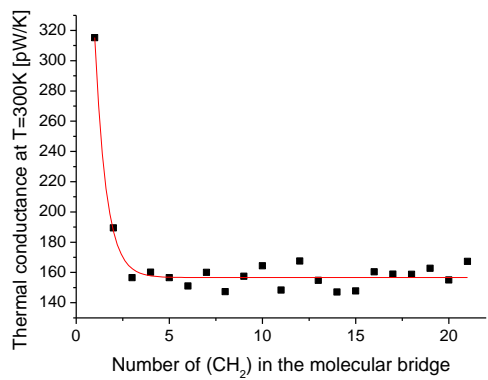

(b)

Figure 2 Thermal conductance vs temperature, for selected chain length (a) and Thermal conductance at 300K as a function of chain length (b). The line in b) denotes the fitting of the data to an exponential decay.

The thermal conductance at room temperature (Figure 2b) clearly decreases with the increase of alkyl chain length between ($CH_2$) and ($CH_2$)$_3$. However, for longer alkyl chains, it is observed that the thermal conductance levels-off around 160 $\frac{pW}{K}$. These results are in accordance with previous theoretical works [31,59,60], NEMD simulations [54] and experimental measurements [61]. In those studies, the thermal conductance was predicted to be independent of the chain length. The Fourier's law of thermal conductance implies a $\frac{1}{N}$ dependence on the length of the chain, meanwhile arguments provided by Segal et al. [59] suggest that the conductance may be length independent for long chains; the actual answer to this effect depends on the density of states and the localization properties of the molecular phonon modes. The interplay between the number of phonon modes at each frequency, their energy transfer capability and the spectral density of the heat reservoirs affect the thermal conductance dependence on the chain length. The analysis of the phonon density of states (DOS) gives an indicator of the efficiency of heat transfer [42]. In order to have an efficient phonon transport, the vibrational states of graphene and the molecular junctions need to be coupled effectively. The overlap of the local DOS of the bridge and the graphene sheets explains the efficiency of the conductance at the junctions. Sasikumar et al. [41] found that the local phonon DOS of a chain connected to a crystalline solid acquires characteristics of the crystal spectrum, particularly in the low frequency range.

The local DOS, the phonon transmission function and the cumulative phonon transmission of selected aliphatic molecular junctions are reported in Figure 3. A significant evolution of both transmission spectra and DOS is observed when the -(CH$_2$)$_n$-chain length increases from n=1 to n=4, especially in the frequency regions 546-820 $cm^{-1}$ for transmission spectra and below 245 $cm^{-1}$ for phonon DOS, corresponding to acoustic modes. There is also an increase of the complexity of the DOS spectra between 750 $cm^{-1}$ and 1500 $cm^{-1}$, which comes from the increasing number of atoms in the chain. On the other hand, for n=5 or higher, minor differences may be observed in the spectra, which explains the thermal conductance for medium and long chains remaining constant. These results are possibly explained by a transition from tunneling-like behaviour to resonant-like phonon modes contributions, as previously proposed by Segal et al. [59]. The cumulative phonon transmission of the alkyl chains (Figure 3c), which shows the number of phonons transmitted under a frequency as the frequency evolves, further highlights the differences in the transmission spectra as the length changes and clearly relates with the conductance plots. Indeed, from n=4 onwards, only minor differences are observed.

### 3.2 Aromatic/aliphatic molecular junctions

Diphenoxyalcane junctions have been recently demonstrated experimentally viable by Bernal et al. [53] for the edge functionalization of graphene nanoplatelets. Therefore, in this section, the conductance and phonon transfer on this type of molecules was studied in detail. In particular, effects of chain length as well as structure and conformation were addressed.

**Effect of chain length**

The effect of the alkyl chain length connecting two phenoxyl groups was studied by varying the number of carbon atoms from ($CH_2$) to ($CH_2$)$_5$, as shown on Figure 4a. A slight increase in conductance was observed for ($CH_2$)$_3$ to ($CH_2$)$_5$ chains between the phenoxyl groups compared to shorter chains. This appears to be a consequence of the modification of the transmission function (Figure 4b) where for n=3, 4, 5 there is an increase at frequencies below ~500 $cm^{-1}$, directly affecting the integrated transmission (Figure S2a), whereas limited differences were found in the DOS (Figure S3). The conductance of the junctions increases with temperature (Figure S2b) as previously observed for alkyl chains. Based on these results, the alkyl chain length does not seem to strongly affect the conductance of aromatic/aliphatic junction, which becomes independent on the number of methylene groups for n≥3, similarly to the case of alkyl junctions reported above. However, lower conductance obtained with n≤2 chains were found, possibly related to a higher constrain imposed by bonding to the aromatic structure, which cannot be effectively relaxed on the short alkylchains.

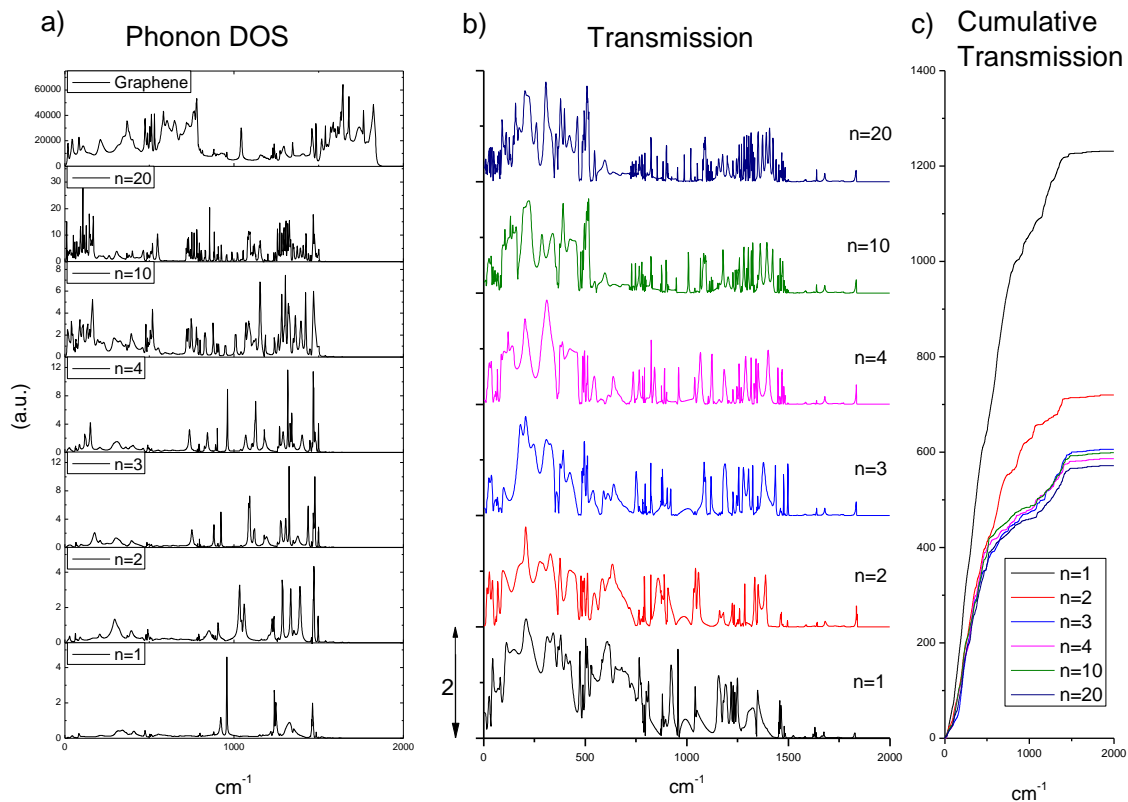

Figure 3 a) Phonon densities of states, b) transmission function spectra and c) cumulative transmission function of $(CH_2)_n$ for n=1,2,3,4,10,20. Transmission function spectra are vertical shifted for clarity.

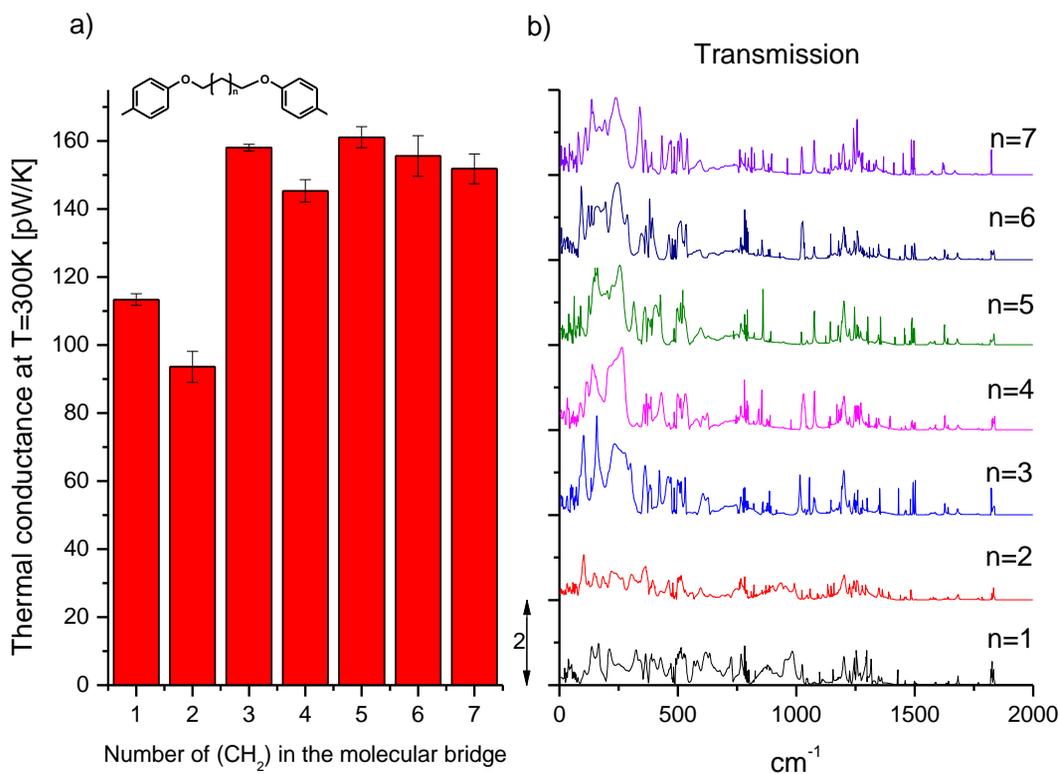

Figure 4 a) Dependence of the thermal conductance with the length of the $(CH_2)_n$ alkyl part of the molecule at T=300K. b) The transmission function. Transmission functions are shifted vertically for clarity.

**Effect of chain conformation**

In this section, diphenoxypentane junctions are further studied for their conformation: in particular, elongated (Figure 5a), compressed (Figure 5b) and out-of-plane (Figure 5c) diphenoxypentane junctions are addressed.

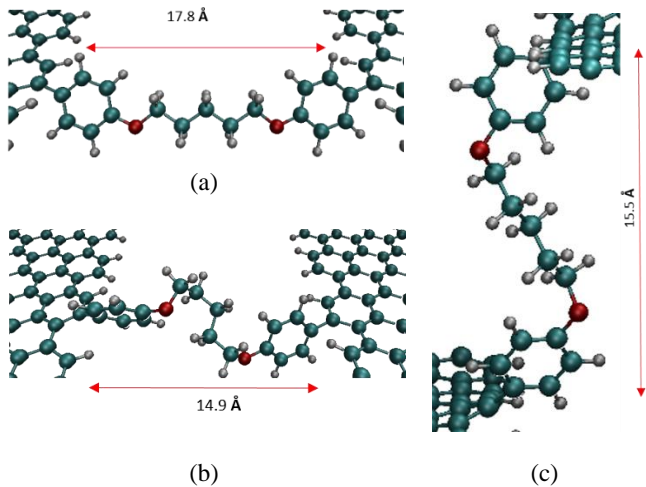

Figure 5: Graphic representation of the elongated (a), compressed (b) and out-of-plane (c) diphenoxypentane molecular junctions.

Besides the fundamental study, this is aimed at obtaining a representative evaluation of thermal conductance for practical applications of molecular junctions, in which the precise control of the molecular conformation is extremely challenging.

In the case of the elongated molecule, the positions of the atoms are initially defined after the relaxation to the local minimum potential energy of the free molecule. After that, the graphene is placed at the distance defined by the elongated molecule and then the whole system is relaxed again at the minimum of potential energy. On the other hand, when the junction is compressed, a number of different chain conformations are possible with the same or similar energies. Therefore, six different conformations of the molecular junction were calculated to average them and obtain a representative conductance range. In the out-of-plane case the graphene contacts are situated on parallel planes at a distance of 15.5Å, selected to keep molecule elongated.

Thermal conductance at T=300K of elongated, compressed, and out-of-plane junctions are $160\pm4\frac{pW}{K}$, $136\pm24\frac{pW}{K}$ and $109\pm9\frac{pW}{K}$, respectively, and their conductance vs. temperature monotonically increasing for temperature varying between 100 K and 500 K (Figure S5). These calculations are in accordance with the results of Li et al. [43], who reported halved conductance for a compressed alkyl junction compared with the elongated one. Furthermore, it is evident that the standard deviation of the thermal conductance on the compressed system is larger, owing to the fact that there are multiple possible different configurations on the compressed molecule, which translates into a bigger variability in the thermal conductance. Conductance values for junctions in different conformations are reported in Table S1.

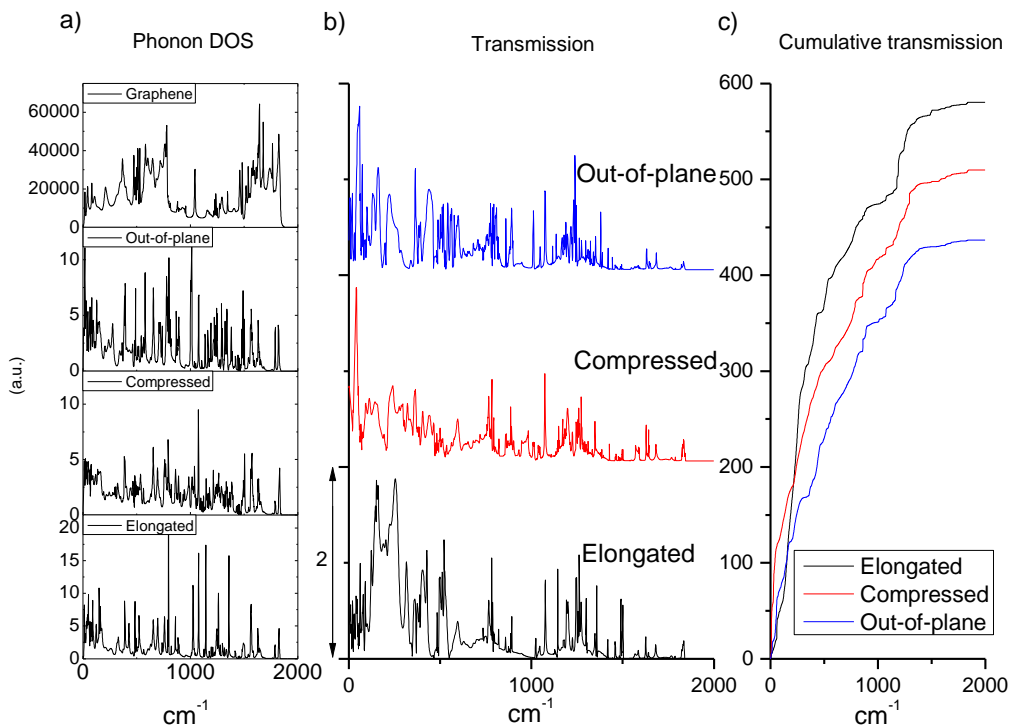

Figure 6 Phonon DOS (a), Transmission function (b) and Cumulative transmission function (c) of elongated, compressed and out-of-plane Diphenoxypentane molecular junctions. The Transmission function are shifted vertically for clarity.

The local phonon density of states in Figure 6a of the elongated junction shows peaks of higher intensity than the compressed and out of plane cases, due to the existence of bends and strains in the latter cases that scatter the states out of those obtained in the elongated junction. The local phonon DOS (Figure 6a) of the compressed junction shows lower intensity at low frequency ($\sim 250\,cm^{-1}$) modes, compared to the elongated counterpart. It was found by Sasikumar et al. [41] that for compressed chains, the chain thermal resistance is proportional to the number of gauche torsional states, demonstrating that these effects dominate phonon scattering. Here, we can see that the phonon DOS of the compressed and out-of-plane molecular junctions have an increased number of peaks at higher frequencies (from $\sim 1150\,cm^{-1}$ to $\sim 1750\,cm^{-1}$, usually identified as the optic modes). This suggests that the phonon spectra contains additional features due to the scattering of those optical vibrational modes as the local orientation of the molecule changes due to deformed bonds or gauche conformations.

The phonon transmission function is also shown in Figure 6b, while the cumulative transmission is shown in Figure 6c. The shape of transmission spectra for the elongated chain is significantly different to the other two cases especially at low frequencies below $\sim 500 cm^{-1}$. In the elongated junction, there is a strong wide peak near 250 $cm^{-1}$ that is displaced into lower frequencies for the compressed and out-of-plane cases, suggesting the main difference in phonon transmission between elongated and non-elongated conformations begins in the acoustic modes.

### 3.3 Aromatic molecular junctions

Aromatic and polyaromatic molecular junctions (Table 1) are expected to constitute more rigid connections that may be exploited to enhance the thermal conductance between graphene sheets, despite graphene edge grafting with fully aromatic molecules is experimentally challenging.

The local phonon DOS, restricted to the molecular junction, are reported in Figure 7. On these spectra, below ~ 800 $cm^{-1}$ we can differentiate between two groups. The first group corresponds to the linear chains (diphenoxybenzene, dibenzylbenzene, diphenylbenzene and biphenyl) which share the same general spectral shape. On the other hand, a second group with phenanthrene and pyrene, shows pronounced peaks at ~ 210 $cm^{-1}$ and double peaks at ~ 775 and ~ 810 $cm^{-1}$. These features are likely related to the higher stiffness of the polyaromatic structures, allowing for out-of-plane vibrations. Significant differences apply also to the transmission function, especially in terms of higher intensity peaks in the range 500–1,200 $cm^{-1}$ for biphenyl, diphenylbenzene, phenantrene and pyrene, which are clearly affecting the cumulative transmission function (Fig. S6 in the Supporting Info)..

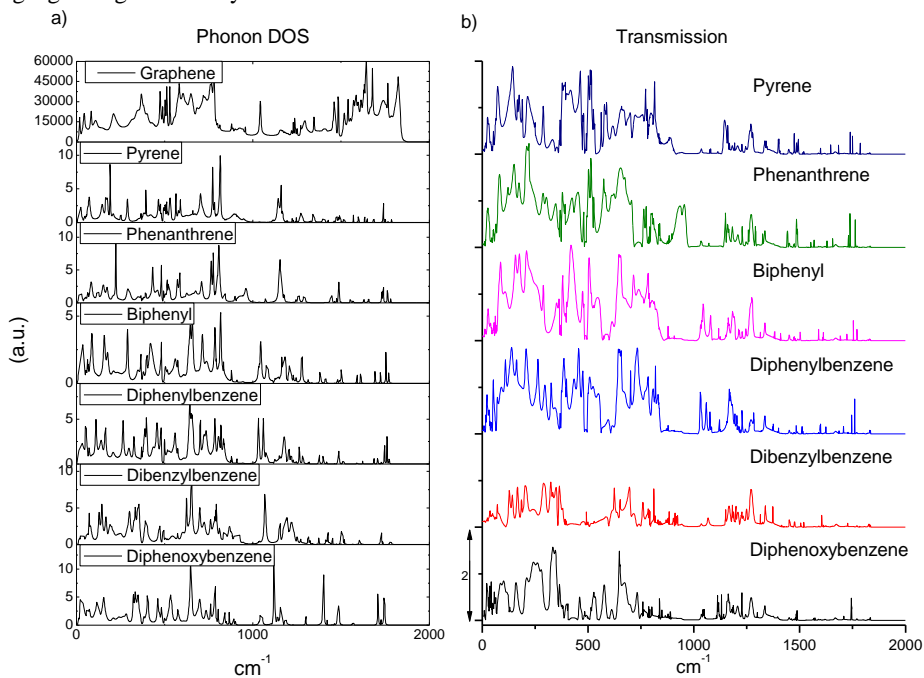

Figure 7: Phonon DOS and transmission function spectra of (from top to bottom): 1,4-diphenoxybenzene, 1,4-dibenzylbenzene, 1,4-diphenylbenzene, biphenyl, phenanthrene and pyrene. The phonon DOS and transmission functions are shifted vertically for clarity.

Thermal conductance values obtained for the different aromatic junctions are reported in Figure 8. The comparison between diphenoxylbenzene to the dibenzylbenzene provides insights on the type of connection between aromatic groups, here either oxygen or methylene. It is worth noting that the differences between inter-benzene linkers are both in the bond angles, with 104° for the oxygen and 109.5° for the ($CH_2$), as well as in the masses , with ~16 u for the oxygen and ~14 u for the ($CH_2$). Despite the effect of different masses [26] and chain branching [27] were previously reported to affect conductance, similar effects are expected to be negligible in our case, as the difference in mass is rather low. Additionally, here the oxygen containing molecule exhibits a higher thermal conductance, whereas one could expect that the mass difference of oxygen compared to carbon could produce an enhanced phonon scattering. However, differences in the stiffness and geometry between C-C-C and C-O-C bonds may explain the difference in the phonon transmissions, that is higher for the diphenoxybenzene especially below 750 $cm^{-1}$ (Figure 7b).

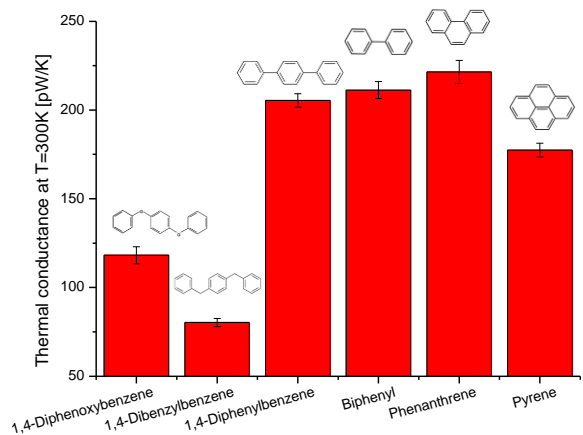

Figure 8 Thermal conductance at T=300K for the aryl molecules addressed.

In order to better investigate this point, we calculated the atom-projected phonon DOS on the oxygen and on the CH$_2$ groups. This analysis, shown in Figure 9, provides information about which frequencies are associated with methylene and oxygen links and allows to correlate to the phonon transmissions.

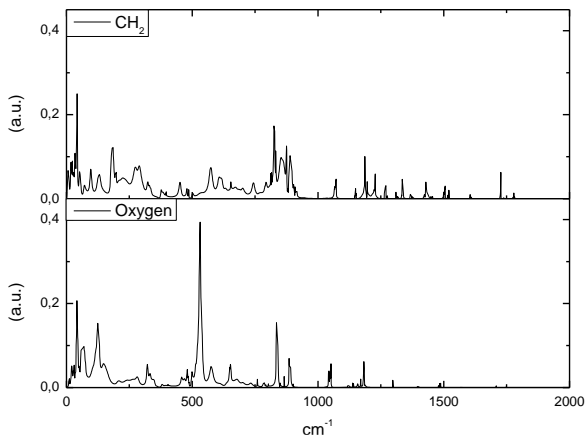

Figure 9 Local DOS of the links between the benzenes of 1,4-diphenoxybenzene and 1,4-dibenzylbenzene.

We found that the DOS on the oxygen links shows sharper and well separated peaks compared to the DOS on the CH$_2$ groups, but these do not correspond to any significant feature in the transmission. On the other hand, the DOS on the CH$_2$ groups is larger, especially in the low frequency range <500 cm$^{-1}$. This may be explained as the diphenyl-benzene has vibrational modes more localized around the CH$_2$ linking groups weakly coupling to the contacts, justifying for the lower transmission seen in Figure 7. On the other hand, by looking in details at the phonon transmissions for dibenzylbenzene and diphenoxybenzene shown in Figure 7b, we observe that in the former case $T(\omega)$ is characterized by several Fano-like features throughout the spectrum. Famili et al. [26] and Klockner et al. [27] discussed how the emergence of phononic Fano interferences could lead to a low thermal conductance in molecular bridges. Fano resonances could arise because the CH$_2$ linkers have masses and therefore vibrational frequencies in resonance with those of the benzene rings, whereas the mass mismatch of the oxygens breaks the mode degeneracies, removing destructive interferences in the transmission.

For the polyaromatic molecular junctions, first we have a decrease of the thermal conductance going from biphenyl to para-terphenyl. This is expected as a similar evolution was seen in the length-dependent transmission of the alkyl chains. A significant increase is also observed for thermal conductance of phenanthrene, compared to biphenyl, as a consequence of the higher stiffness of polycondensed aromatics due to the increase of C-C bonds. However, the further aromatic condensation in pyrene does not lead to a further increase of thermal conductance, that rather unexpectedly shows a significant step down. We observe that the phonon DOS does not show a significant change, however the transmission function of phenanthrene is slightly larger. This unexpected result may be related to the mismatch between the plane of the molecule and the plane of graphene sheets. Indeed, in our calculations, the biphenyl, duphenylbenzene and pyrene molecular bridges are found to relax into a configuration with their planes perpendicular to the plane of the graphene contacts (Figure S6). This may affect the transmission of graphene in-plane phonon modes perpendicular to the plane of the molecule, thus causing molecule in-plane stiffening to be ineffective for thermal transfer between the graphene flakes.

## 4 Conclusions

A combination of the atomistic DFTB method with a Green's functions formalism was applied in this paper, in order to calculate the thermal conductance of several aliphatic, aliphatic-aromatic and aromatic molecular junctions.

The variation of the thermal conductance with the chemical structure and conformation of the molecular bridge evidenced the possibility to engineer contacts between graphene nanoflakes and, hence, to control the thermal transport in graphene-based nanostructures.

In particular, the evolution of the thermal conductance as a function of the length on alkyl molecular junctions has been investigated, showing extremely high conductance values for methylene, while a sudden asymptotic drop was obtained for longer alkyl chains, leading to values in the range of 150 pW/K for -(CH$_2$)$_3$- and longer chains. Beside the effect of chain length, dependency of thermal conductance on chain conformation (elongated, compressed, and out-of-plane) was demonstrated and explained in terms of scattering of phonons on strained bonds and gauche conformations. Unfortunately, the experimental manufacturing of alkyl chain molecular junctions remains extremely challenging, especially by short bridges (e.g., methylene, ethylene). More experimentally viable aliphatic-aromatic junctions were also studied, correlating conductance, phonon density of states and phonon transmission to their chemical structure. The presence of structurally rigid aromatic rings in the molecular junction was found to be generally beneficial in terms of higher thermal conductance for a given length of the junction. Furthermore, conductance and phonon transmission spectra were found to strictly depend on the nature of the groups connecting aromatic rings. Indeed, differences were found between ether and methylene bridges connecting phenylenes, attributed to a suppression of phonon transmission due to destructive Fano interferences. Fully aromatic structures were found to deliver the highest thermal conductance values (>200 pW/K) explained by their molecular stiffness, resulting in high phonon transmission, especially in the frequencies below 1200 cm$^{-1}$.

The results obtained in this paper by the screening of several different types of molecular junction structure will provide a theoretical basis for the design of controlled thermal interfaces to be exploited in engineered thermally efficient nano-devices.

## Acknowledgements

This work has received funding from the European Research Council (ERC) under the European Union's Horizon 2020 research and innovation programme grant agreement 639495 — INTHERM — ERC-2014-STG. L.M.S. thanks the Deutscher Akademischer Austauschdienst (DAAD) for the financial support. This work has also been partly supported by the German Research Foundation (DFG) within the Cluster of Excellence "Center for Advancing Electronics Dresden". B.M. greatly acknowledges the financial support by European Research Council for COMBAT project (Grant number 615132).

Diego Martinez Gutierrez[1], Alessandro di Pierro[1], Alessandro Pecchia[2], Leonardo Medrano Sandonas[3,4], Rafael Gutierrez[3], Mar Bernal[1], Bohayra Mortazavi[5], Gianaurelio Cuniberti[3,4,6], Guido Saracco[1], Alberto Fina[1] (✉)

*1-Dipartimento di Scienza Applicata e Tecnologia, Politecnico di Torino, 15121 Alessandria, Italy*
*2-Consiglio Nazionale delle Ricerche, ISMN, 00017 Monterotondo, Italy*
*3-Institute for Materials Science and Max Bergmann Center of Biomaterials, TU Dresden, 01062 Dresden, Germany*
*4-Center for Advancing Electronics Dresden, TU Dresden, 01062 Dresden, Germany*
*5- Institute of Structural Mechanics, Bauhaus-Universität Weimar, D-99423 Weimar, Germany*
*6- Dresden Center for Computational Materials Science, TU Dresden, 01062 Dresden, Germany.*

## Theoretical background

The bulk heat conductivity of a material with a moderate temperature gradient is related to the heat flux via the Fourier equation

$$\frac{1}{A}\frac{dQ}{dt} = -\kappa_{bulk}\frac{dT}{dr} \qquad (1)$$

where $\frac{dQ}{dt}$ represents the amount of heat transferred per unit time along the gradient, A is the cross-section perpendicular to the heat flux direction. So the heat conductivity is

$$-\kappa_{bulk} = \frac{1}{A\frac{dT}{dr}}\frac{dQ}{dt} \qquad (2)$$

The presence of an interface or defects between the temperature source and sink implies a discontinuity in the temperature profile as these elements provide additional thermal resistance. In those situations the interfacial conductance is defined as

$$G = \frac{1}{R_i} = \frac{1}{A\Delta T}\frac{dQ}{dt} \qquad (3)$$

In the case of a molecular junction, the interfacial conductance per chain can be defined as

$$G = \frac{1}{R_i} = \frac{1}{\Delta T}\frac{dQ}{dt} \qquad (4)$$

As the definition of temperature at the nanoscale is not trivial, usually temperatures are defined from the kinetic energy as

$$\frac{3}{2}Nk_BT = \frac{1}{2}\sum_{i=1}^{N}m_iv_i^2 \rightarrow T = \frac{1}{3Nk_B}\sum_{i=1}^{N}m_iv_i^2 \qquad (5)$$

where N is the number of atoms and the 3 factor comes from the usual three degrees of freedom. A more rigorous definition of temperature is shown by S. Lepri et al. [1].

## Density Functional Tight Binding

Tight-binding is a method for modelling band-structures with one to several fitted hopping parameters. Density functional tight binding uses DFT in order to obtain the parametrizations, thus this makes the DFTB method deeply rooted to first principles. But DFTB is not an ab-initio method since it contains parameters. Most of those parameters have a solid theoretical basis.

For a full derivation of the DFTB method see Foulkes et al. [2], Frauenhein et al. [3] and Elstener et al. [4]. For a more didactic approach of DFTB see Koskinen and Maakinen [5].

Basically it begins with the Kohn-Sham DFT [6] energy:

$$E[n] = \sum_a f_a <\psi_a|(-\frac{1}{2}\nabla^2 + \int V_{ext}(\vec{r})n(\vec{r})d^3\vec{r})|\psi_a> + \\ + \frac{1}{2}\int\int \frac{n(\vec{r})n(\vec{r'})}{|\vec{r}-\vec{r'}|}d^3\vec{r}d^3\vec{r'} + E_{xc}[n] + E_{II} \qquad (6)$$

After some algebra, we can describe the energy as

$$E[\delta n] = E_{BS}[\delta n] + E_{Coul}[\delta n] + E_{rep} \qquad (7)$$

And then we apply the Tight-Binding formalism by considering the matrix elements of the Hamiltonian as parameters, thus:

$$\begin{aligned}
E = &\sum_a f_a \sum_{\mu\nu} c_\mu^{a*} c_\nu^a H_{\mu\nu}^0 &&\Leftarrow B.S. \\
&+ \tfrac{1}{2}\sum_{ij} \gamma_{ij}(R_{ij}) \Delta q_i \Delta q_j &&\Leftarrow E_{Coul.} \\
&+ \sum_{i<j} V_{rep}^{ij}(R_{ij}) &&\Leftarrow E_{rep.}
\end{aligned} \qquad (8)$$

In this approximation, the atomic forces can be calculated directly from the final expression of the DFTB energy (eq.8) by the use of gradients with respect to the coordinates $R_{ij}$ ($\nabla_j = \frac{\partial}{\partial R_j}$)

$$\begin{aligned}
F_i = &-\sum_a f_a \sum_{\mu\nu} c_\mu^{a*} c_\nu^a [\nabla_i H_{\mu\nu}^0 - (\varepsilon_a - h_{\mu\nu}^1)\nabla_i s_{\mu\nu}] - \\
&- \Delta q_i \sum_j (\nabla_j \gamma_{ij}) \Delta q_j - \nabla_i E_{rep}
\end{aligned} \qquad (9)$$

the $\nabla_i \gamma_{ij}$ gradients are obtained analytically from the Coulombic energy equation $E_{Coul} = \frac{1}{2} \sum_{ij} \gamma_{ij(r_{ij})} \Delta q_i \Delta q_j$. $\nabla_i H_{\mu\nu}^0$ and $s_{\mu\nu}$ are calculated numerically with interpolation (Slater-Koster integrals) and use to be tabulated.

## Green's functions formalism

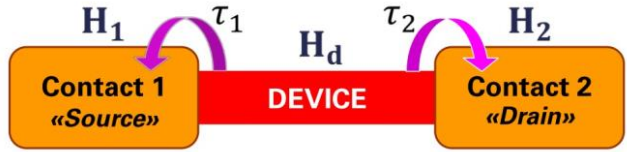

Figure S1: Schematic of the typical phonon transport problem considered in the Atomistic Green's Function formulation.

The Green's function formalism is based on the study of the harmonic matrix between contacts/leads and elements [7]. Harmonic matrix [8]:

$$H = \{H_{p,q}\} = \frac{1}{\sqrt{M_p M_q}} \begin{cases} \dfrac{\partial^2 U}{\partial u_p \partial u_q} & \text{if } p \neq q \\ -\sum_{k \neq q} \dfrac{\partial^2 U}{\partial u_p \partial u_k} & \text{if } p = q \end{cases} \qquad (10)$$

where up and uq refer to any two atomic vibrational degrees of freedom (i.e., displacements), respectively. U represents the total interatomic potential. Mp and Mq are atomic masses associated with degrees of freedom up and uq, respectively.

The dynamical equation can be written as:

$$[\omega^2 I - H]\Phi = 0 \qquad (11)$$

Where $\Phi_p = u_p \sqrt{M_p}$ (M is the atomic mass).

From here we can obtain the Green's functions [8], [9]

$$G_d = [\omega^2 I - H_d - \Sigma_1 - \Sigma_2]^{-1} \qquad (12)$$

Where $G_d$ is the device subset of the overall Green's Function, and

$$\begin{aligned} \Sigma_1 &= \tau_1 g_1 \tau_1^\dagger \\ \Sigma_2 &= \tau_2 g_2 \tau_2^\dagger \end{aligned} \qquad (13)$$

where τ are the connection matrix between the contacts and the device, and g are the uncoupled Green's functions of the contacts.

From here, we can obtain the Densities of states using the fact that the LDOS is given by the diagonal components of the spectral function of the isolated contact

$$LDOS(E) \sim Im(Tr[G(E)]) \qquad (14)$$

Also we can calculate the heat flow from/to the contact to the device

$$J_1 = \int_0^\infty \frac{\hbar\omega}{2\pi} Tr[\Gamma_1 \underbrace{G_d \Gamma_2 G_d^\dagger}][f_{BE}^o(\omega,T_1) - f_{BE}^o(\omega,T_2)]d\omega \qquad (15)$$

where $\Gamma_1 = i[\Sigma_1 - \Sigma_1^\dagger] = i[\tau_1(g_1 - g_1^\dagger)\tau_1^\dagger] = \tau_1 a_1 \tau_1^\dagger$ is the "escape rate" and $f_{BE}^o$ is the Bose-Einstein distribution.

And from here, the thermal conductance can be calculated

$$\mathcal{G} = \int_0^\infty \frac{\hbar\omega}{2\pi} Tr[\Gamma_1 G_d \Gamma_2 G_d^\dagger] \frac{\partial f_{BE}^o}{\partial T} d\omega \qquad (16)$$

and the transmission

$$\mathcal{T}(\omega) = Tr[\Gamma_1 G_d \Gamma_2 G_d^\dagger] \qquad (17)$$

## Additional graphs and tables

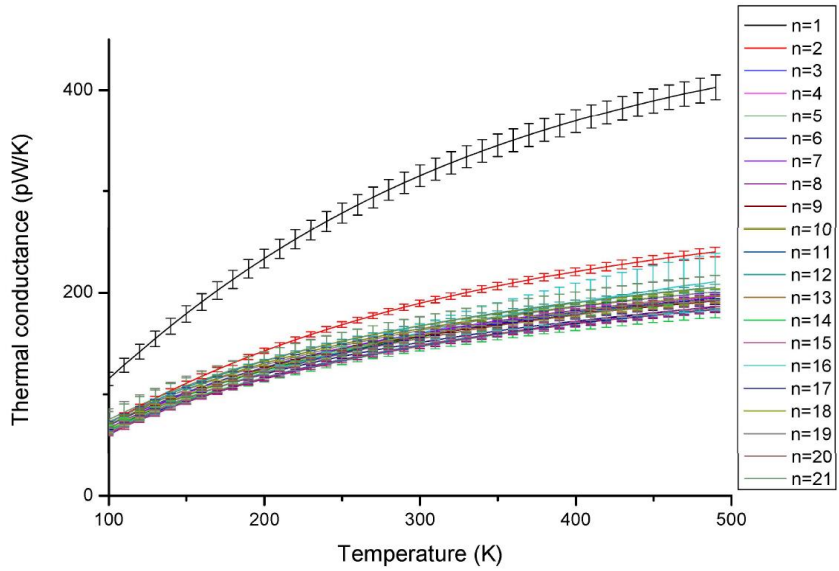

Figure S2 Thermal conductance vs. temperature for all calculated alkyl chains.

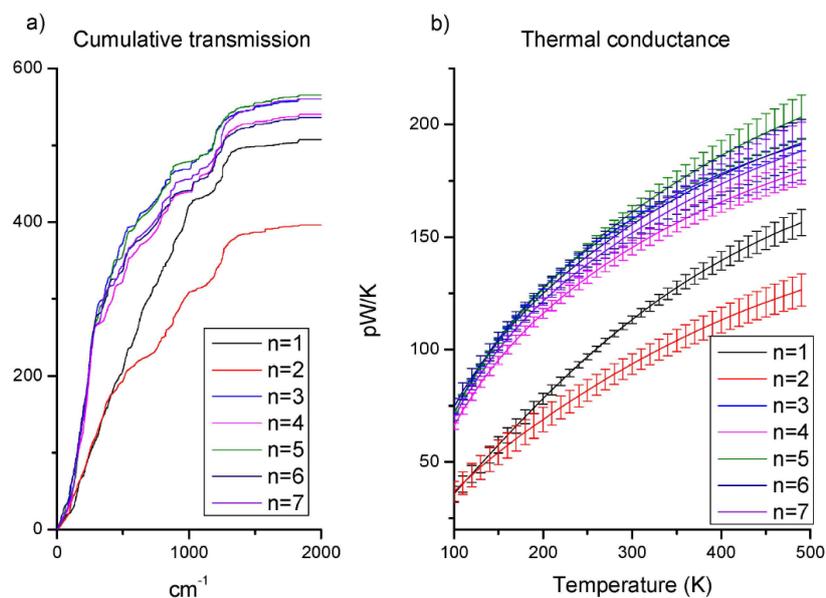

Figure S3 Cumulative transmission function (a) and thermal conductance (b) evolution with the length of the $(CH_2)_n$ alkyl part of the alkyl-aryl molecular junction.

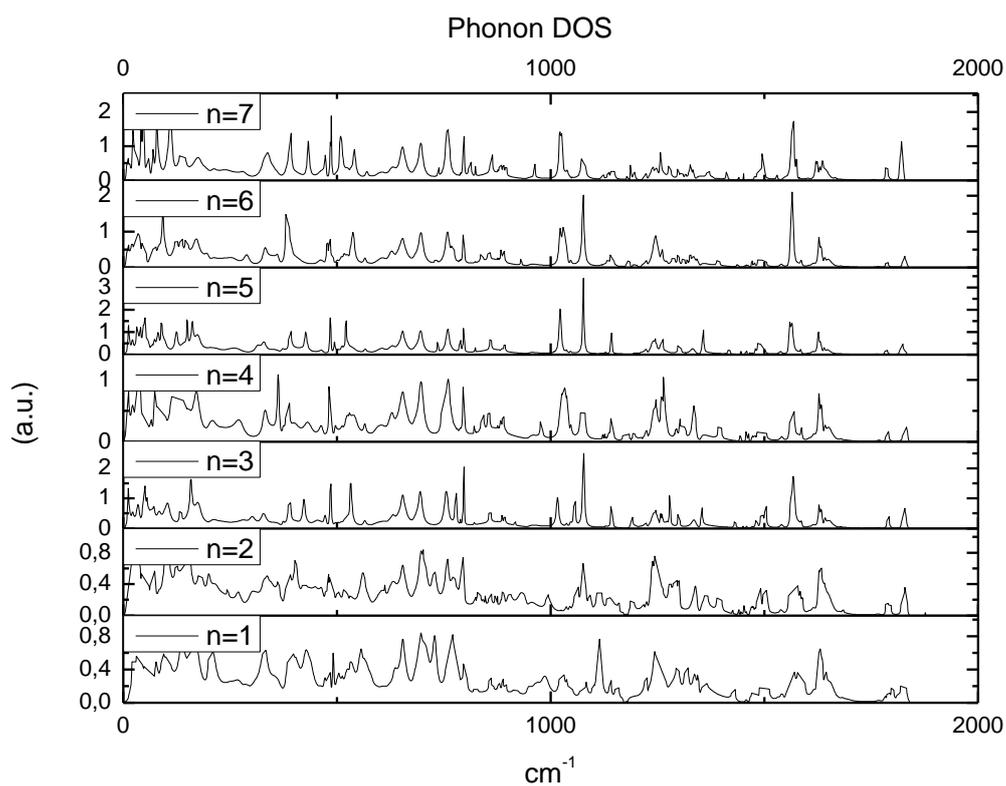

Figure S4 Phonon DOS of the alkyl chain length connecting two phenoxyl groups in extended conformation

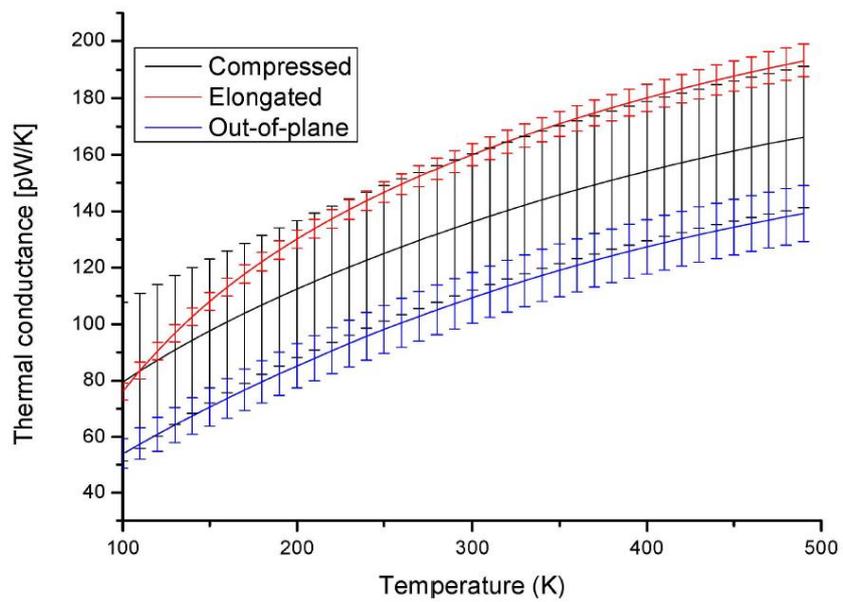

Figure S5 Thermal conductance vs. temperature of the different studied conformations of diphenoxypentane

Table S1 Conductance values (pW/K) for junctions in different conformations at T=300K.

| | Run 1 | Run 2 | Run 3 | Run 4 | Run 5 | Run 6 | Average | St. dev. |
|---|---|---|---|---|---|---|---|---|
| Elongated | 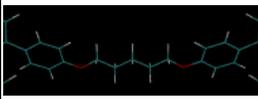 156 | 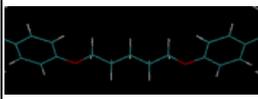 162 | 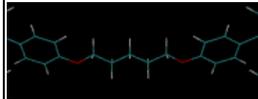 155 | 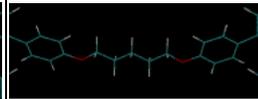 163 | 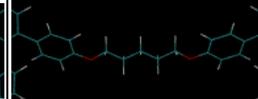 159 | 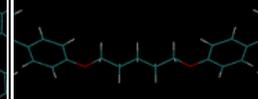 166 | 160 | 4 |
| Compressed | 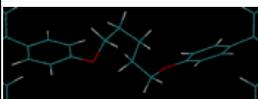 115 | 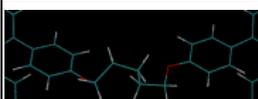 161 | 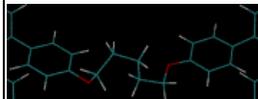 172 | 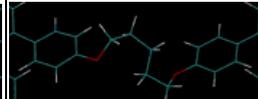 103 | 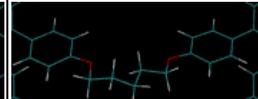 127 | 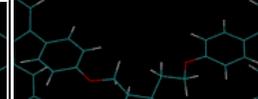 136 | 136 | 24 |
| Out of Plane | 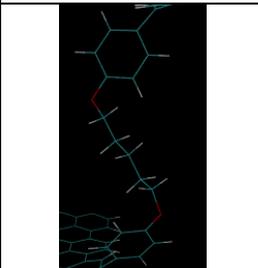 125 | 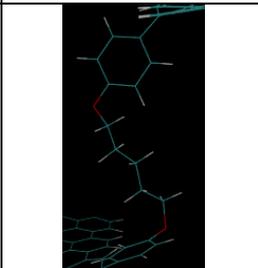 115 | 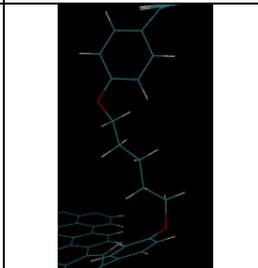 100 | 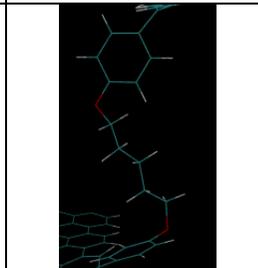 106 | 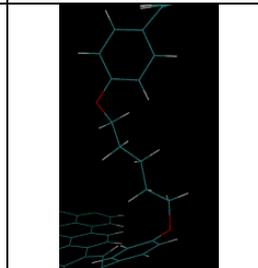 99 | 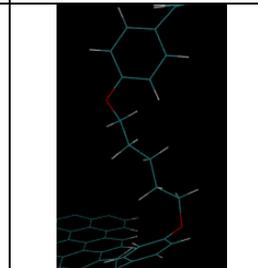 112 | 109 | 9 |

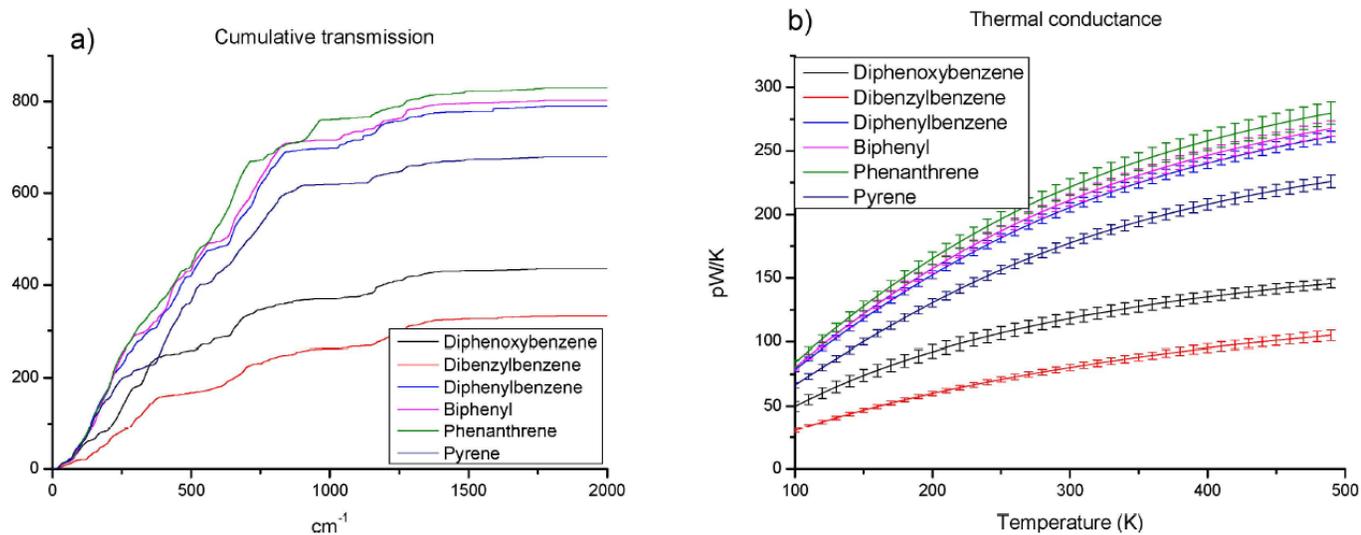

Figure S6 Cumulative transmission function (a) and thermal conductance (b) of 1,4-Diphenoxybenzene, 1,4-Dibenzylbenzene, 1,4-diphenylbenzene, Biphenyl, Phenanthrene and Pyrene

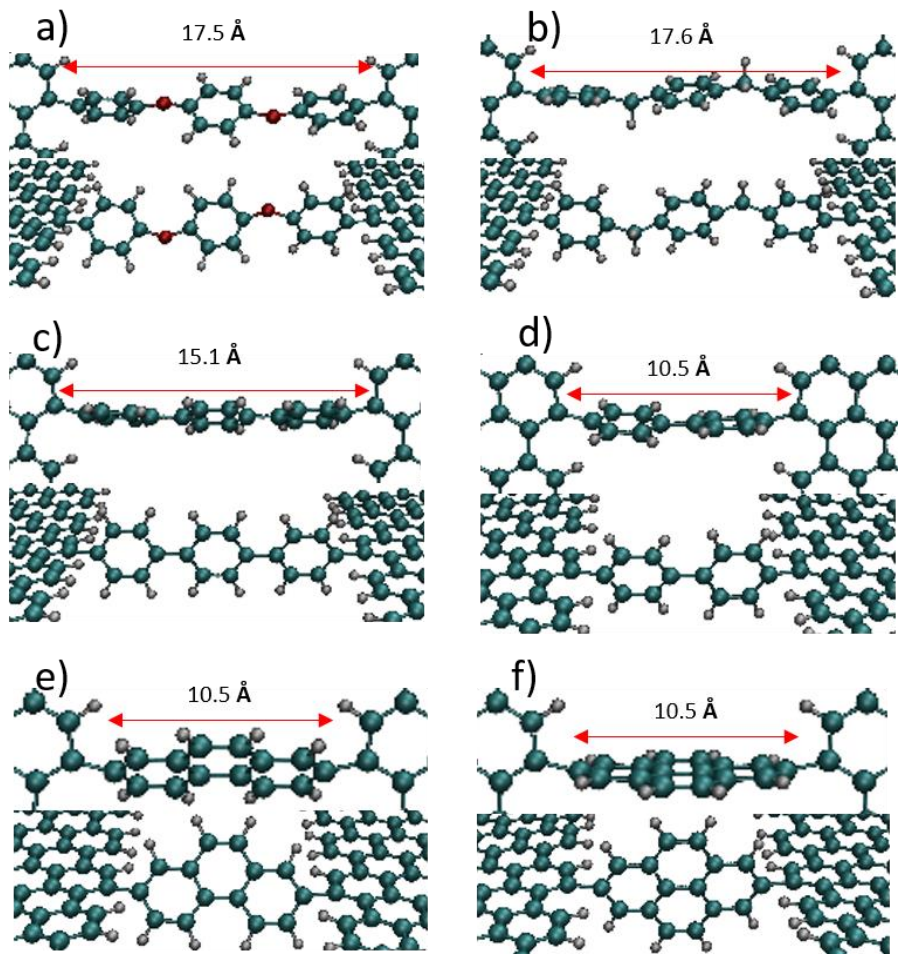

Figure S7 Aromatic molecular junctions in their conformation assumed after relaxation to minimize potential energy (Top and 45° views): 1,4-diphenoxybenzene (a), 1,4- dibenzylbenzene (b), 1,4-diphenyl benzene (c), biphenyl (d), phenanthrene (e) and pyrene (f).

Address correspondence to Alberto Fina, alberto.fina@polito.it